\documentclass[preprint,5p]{elsarticle}

\usepackage{amsmath,amssymb,amsfonts}
\usepackage{mathtools}
\usepackage{cleveref}
\usepackage{siunitx}

\usepackage{xcolor}

\newcommand{\RR}{\mathbb{R}}
\newcommand{\ZZ}{\mathbb{Z}}
\newcommand{\eps}{\varepsilon}
\DeclareMathOperator*{\eE}{\mathbb{E}}

\begin{document}

\begin{frontmatter}
    \title{The WQN algorithm for EEG artifact removal in the absence of scale invariance
        \tnoteref{funding}}
    \author[1]{Matteo Dora}
    \ead{matteo.dora@ieee.org}
    
    \author[2]{Stéphane Jaffard}
    \ead{jaffard@u-pec.fr}
    
    \author[1]{David Holcman \corref{corrauth}}
    \ead{david.holcman@ens.psl.eu}
    \cortext[corrauth]{Corresponding author: david.holcman@ens.psl.eu.}
    
    \affiliation[1]{%
        organization={IBENS UMR8197, École Normale Supérieure -- PSL},
        postcode={75005},
        city={Paris},
        country={France}
    }
    \affiliation[2]{%
        organization={Univ Paris Est Creteil, Univ Gustave Eiffel, CNRS, LAMA UMR8050, F-94010 Creteil, France},
        postcode={F-94010},
        city={Créteil},
        country={France}
    }
    \tnotetext[funding]{Funding: D.H.'s research is supported by grants ANR NEUC-0001, PSL and CNRS pre-maturation, and by the European Research Council (ERC) under the European Union's Horizon 2020 research and innovation program (grant agreement Nº~882673).}
    
    \begin{abstract}
        Electroencephalogram (EEG) signals reflect brain activity across different brain states, characterized by distinct frequency distributions. Through multifractal analysis tools, we investigate the scaling behaviour of different classes of EEG signals and artifacts. We show that  brain states associated to sleep and general anaesthesia  are not in general characterized by scale invariance. The lack of scale invariance motivates the development of artifact removal algorithms capable of operating independently at each scale. We examine here the properties of the wavelet quantile normalization algorithm, a recently introduced adaptive method for real-time correction of transient artifacts in EEG signals.
        We establish general results regarding the regularization properties of the WQN algorithm, showing how it can eliminate singularities introduced by artefacts, and we compare  it to traditional thresholding algorithms. Furthermore, we show that the algorithm performance is independent of the wavelet basis. We finally examine its continuity and boundedness properties and illustrate its distinctive non-local action on the wavelet coefficients through pathological examples.
    \end{abstract}
\end{frontmatter}

\section{Introduction}

\noindent For nearly a century, brain activity has been measured using electroencephalography (EEG), a technique that uses electrodes placed on the scalp of a patient to record the electrical activity of the brain~\cite{schomer2012niedermeyer}. This physiological signal reflects the collective activity of neuronal populations~\cite{kandel2000principles}. By analyzing the statistical and spectral properties \cite{le2007analysis} of the EEG, it is possible to identify transient brain oscillations within a frequency range between zero and hundreds of Hz that reflects key cognitive events, such as specific responses to sensory stimulations, learning and memory, sleep stages, meditation, coma, and more.

Because of its non-invasive nature, EEG recordings have been widely  adopted in the clinical setting for screening or monitoring tasks. An example is general anaesthesia (GA), a procedure consisting in placing the brain into an artificial but reversible coma state, which can now be routinely monitored in real-time during surgery by recording the EEG signal from few electrodes. This setting provides a continuous feedback about the depth of anaesthesia, allowing for accurate control of the anaesthetic dose required to keep the patient in a safe unconscious state.

However, the amplitude of the EEG signal typically varies in the microvolt range, making the EEG highly susceptible to contamination by artifacts originating from several types of sources. The issue is particularly pronounced in the clinical setting, where there is a limited control over the environment and the potential for artifact contamination. Artifacts include noise from electrical equipments, muscle contractions, eye movements, or small displacements of the electrodes, that alter the EEG signal.

Eliminating artifacts from the EEG signal is thus a crucial concern. Wavelet-based methods ~\cite{flandrin2013time,donoho1994ideal,johnstone1997wavelet} have been used extensively in this regard by thresholding coefficients to remove artifacts from the EEG signal, taking advantage of the different properties of the artifact and the physiological EEG~\cite{krishnaveni2006removal,inuso2007waveletica,chavez2018surrogate} distributions. In this direction, we recently introduced an empirical method, the WQN algorithm~\cite{dora2022adaptive,dora2022wqn}, designed to remove artifacts from single-channel EEGs for real-time applications in clinical monitoring. This adaptive approach allows attenuating transient artifacts in the EEG by normalizing the wavelet coefficient distribution during the artifact so that it matches the one obtained from a preceding uncontaminated interval.

While our previous studies have demonstrated the high effectiveness of the WQN algorithm in removing transient artifacts from EEG~\cite{dora2022adaptive,dora2022wqn}, the underlying reasons for its efficiency remain unclear. In the current article, we provide a comprehensive analysis of the WQN algorithm in two directions: First, we study the statistical properties of the EEG signal, and second, we derive several properties of the WQN algorithm.

The manuscript is organized as follows. In Section~\ref{sec:multifractal}, we use  multifractal analysis techniques~\cite{Jaf9,jaffard2001wavelets,CiuCiuetal} in order to characterize the scaling behaviour of different classes of EEG signals and artifacts, showing that scale-invariance is not always observed. This result justifies the need to modify wavelet coefficients separately for each scale, as proposed by the WQN algorithm. In Section~\ref{sec:properties}, we describe the WQN algorithm and study its properties. In particular, we show that the WQN algorithm cannot introduce unwanted singularities in the signal. We then show that the WQN algorithm is robust with respect to changes of the wavelet basis. We also put in evidence  its non-local action on the wavelet coefficients and explore some of its consequences;  finally we demonstrate how it performs with respect to pathological cases where signals are perturbed by different types of singularities and random noise.

\section{Scaling properties of the EEG signal}\label{sec:multifractal}
\noindent In the spectral analysis of the EEG signal, the scale invariance property is characterized by a power-law decay present in the power spectrum 
\begin{equation}\label{eq:spectral_inv}
S(f) \sim 1/f^{a},
\end{equation}
with exponent $a$. A scale invariant behavior can provide valuable insights into the underlying mechanisms of neuronal networks \cite{Tsodyks1997}. In this section we investigate in which contexts the EEG signal shows such scale-invariant behaviour. We recall that power spectrum of the activity of local neuronal ensembles can be fitted by a power law~\cite{buzsaki2014log}, however the EEG signal often contains additional oscillatory activity revealed by the presence of specific brain waves, such as the $\alpha$ wave in the range 8–12~Hz, which cannot be filtered without significantly altering the properties of the signal.
\subsection{Wavelet multifractal analysis framework}
\noindent To estimate the scaling properties of the EEG, we resort to the setting supplied  by multifractal analysis~\cite{jaffard2001wavelets,Bergou,CRAS2019}, which extends the notion of scale invariance to signals that cannot be characterized by a single scaling exponent; this approach allows to  overcome the limitation to second-order statistics of the power spectrum by replacing the Fourier transform with a multiresolution tool such as the wavelet transform. Practically, in the wavelet-based multifractal framework, the scale invariance of a signal $x(t)$ is measured through a \textit{wavelet scaling function} $\eta_x(q)$, which encapsulates the scaling exponents associated with the moments of order $q$ ($q >0$) of the wavelet coefficients. This approach can be seen as a generalization of traditional spectral scale-invariance analysis (as in \cref{eq:spectral_inv}), which is recovered for $q = 2$~\cite{CRAS2019}.

To define the scaling function $\eta_x$, we start with the wavelet decomposition for functions defined on $\RR$ \cite{daubechies1992ten,Mey90I,mallat1989theory,jaffard2001wavelets}. We use a { \em mother wavelet}  $\psi$, such that an orthonormal basis of $L^2(\RR)$ is obtained by dilations and translations of $\psi$:
\begin{equation}
    \left\{ \psi_{j, k}\left(t\right) = 2^{-j/ 2} \, \psi\left(2^{-j}\, t - k\right),\, j, k \in \ZZ\right\}.
\end{equation}
For a given  signal $x(t)$, its decomposition on the wavelet basis is given by
\begin{equation}\label{eq:signal_dwt}
    x(t) = \sum_{j,k \in \ZZ} c_{j,k} \, \psi_{j,k}(t),
\end{equation}
where $c_{j,k} = \left< x, \psi_{j,k} \right>$ are the  wavelet coefficients\footnote{The scalar product is defined by $\left< f, g \right> = \int_\RR  f(t) \, g(t) dt $.} of $x$ at scale $j$. We define the \textit{wavelet structure functions} of $x$ based on its normalized wavelet coefficients, as follows:
\begin{equation}\label{eq:structure_function}
    \forall q >0, \qquad S_x (j, q) = 2^j\sum_k \left|2^{-j/2} c_{j,k}\right|^q,
\end{equation}
and finally the \textit{wavelet scaling function} $\eta_x(q)$ is implicitly defined by
\begin{equation}\label{eq:scaling_fit}
   \forall q>0,  \qquad  S_x (j, q) \sim 2^{\eta_x(q) j},
\end{equation}
in the limit of small scales, i.e. when $j \rightarrow -\infty$. The mathematical interpretation is
\[ \eta_x (q) = \liminf_{ j\rightarrow -\infty} \frac{ \log_2 (S_x (j, q))}{j} .\]
This quantity provides information about the global regularity of  $x$.  The particular value $q = 2$ allows to recover the information supplied by the traditional spectral analysis~\cite{CRAS2019}, i.e. the { \em Hurst exponent} $\alpha = \eta_x(2) + 1$ for \cref{eq:spectral_inv}. Recall that the Sobolev spaces $H^{s,q} (\RR )$ are composed of functions whose fractional derivatives belong to $L^q (\RR ) $;  $\eta_x $ has the following function space interpretation:
\[
\eta_x (q) = q \cdot \sup\{ s : f \in H^{s,q}(\RR )  \}.
\]
In practice, we shall compute the wavelet scaling function $\eta_x$ using a log-log plot regression, i.e. fitting $ \log_2 S_x (j, q)$ as a linear function of $j$, in the range of scales where the EEG signal is most informative (typically in the range 0.1–100~Hz). Note that the multifractal framework is pertinent only if the log-log plot is approximately linear in a sufficiently large range of scales, i.e. if the data exhibit an average scaling invariance. We will examine below whether this holds or not for different types of EEG signals.

Another relevant parameter which has proved useful in the setting of  scale invariance is the \textit{uniform Hölder exponent} $H^\text{min}_x$ which can be defined as follows. The largest normalized wavelet exponent at scale $j$ is computed as
\[ D_x (j) = \sup_k 2^{-j/2} |c_{j,k}| ,  \]
then the scaling exponent  $H^\text{min}_x$ is implicitly defined by
\[ D_x (j)  \sim 2^{H^\text{min}_x j}  \quad \mbox{ when } j \rightarrow -\infty ,  \]
i.e.
\[ H^\text{min}_x = \liminf_{ j\rightarrow -\infty} \frac{ \log_2 (D_x (j) )}{j} .\]
Its function space interpretation is obtained through the H\"older spaces $C^s  (\RR )$:
\[ H^\text{min}_x =  \sup\{ s : x \in C^{s}(\RR )  \} . \]
Similarly to the scaling function, the coefficient $H^\text{min}_x$ can be estimated by log-log regression of $D_x(j)$ versus $2^j$. As above, this approach is only applicable when the log-log plot is well described by a straight line on a relevant range of scales.
\subsection{Multifractal analysis of the EEG signal}
\begin{figure*}
    \includegraphics[width=\linewidth]{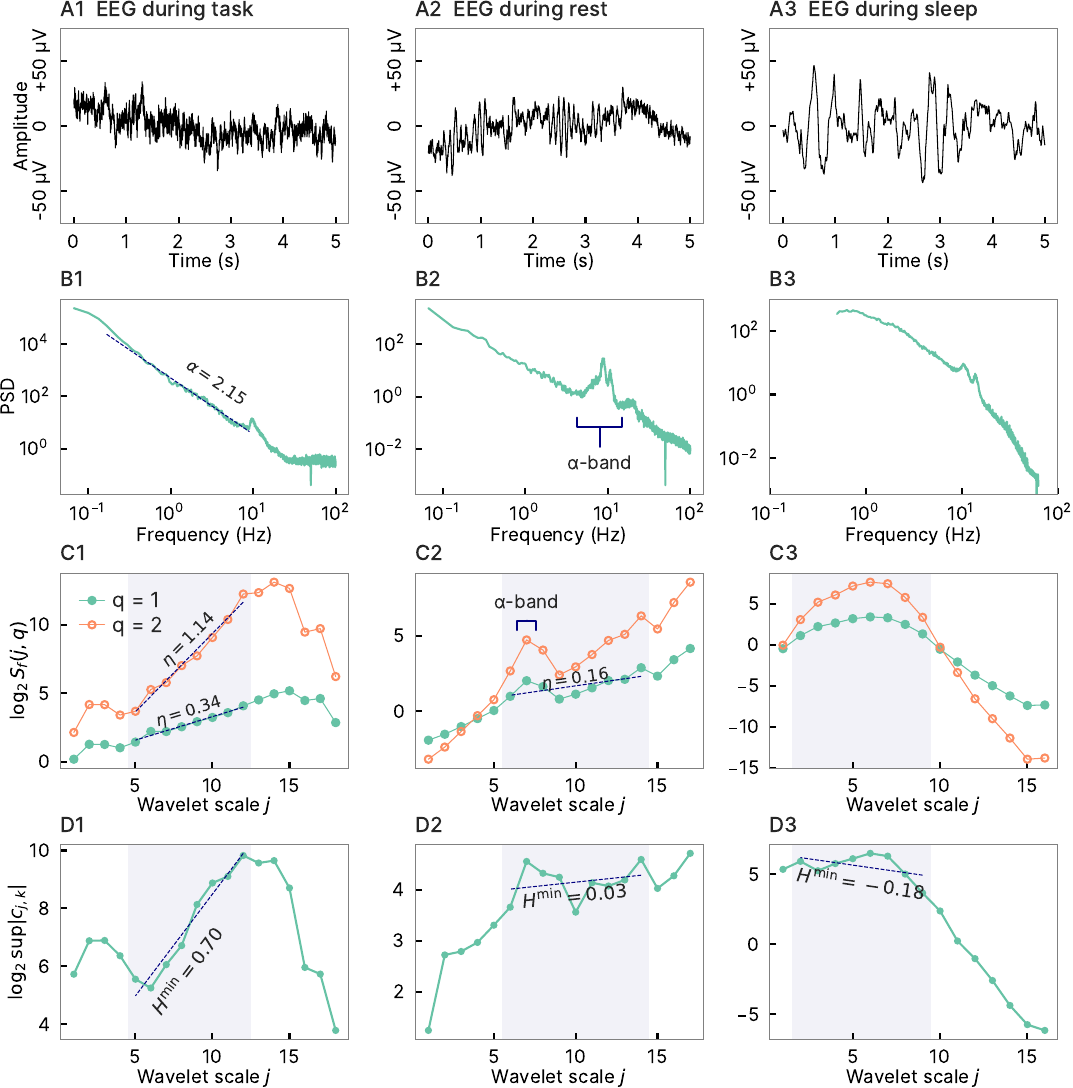}
    \caption{
        \textbf{Multifractal analysis of the EEG signal.}\,
        \textbf{A.}~Samples of EEG signals recorded during task execution (A1), rest (A2), and sleep (A3).
        \textbf{B.}~Power spectral density plots for the three cases (log-log).
        \textbf{C.}~Wavelet structure functions for $q =1, 2$, as described in \cref{eq:structure_function}. The shaded area indicates the range of scales corresponding to a frequency range \SIrange{0.1}{50}{\hertz}. The dashed lines indicate the linear fit that can be used to estimate the scaling function (see \cref{eq:scaling_fit}).
        \textbf{D.}~Estimations of uniform H\"older exponent $H^\text{min}$. EEG during task execution shows a good fit (D1), while rest and sleep give unreliable results (D2–3).
    }\label{fig:mfa}
\end{figure*}

\noindent We now investigate the scaling properties of EEG signals under this multifractal framework. To test whether EEGs present a general scale-invariant structure, we analyze signals acquired in different conditions:
\begin{itemize}
    \item EEG recorded during active task execution \cite{blankertz2007non};
    \item EEG for subjects at rest, with eyes closed \cite{sweeney2012methodology,goldberger2000physiobank};
    \item EEG recorded during sleep \cite{quan1997sleep,goldberger2000physiobank};
\end{itemize}
In general, the high frequency content of EEGs is higher during mental activity and decreases with rest and sleep, as slow and more regular patterns progressively emerge. We visualize this tendency in figure~\ref{fig:mfa}, where we present a comparison of three EEG signals: during task (\cref{fig:mfa}-A1), rest (\cref{fig:mfa}-A2), and sleep (\cref{fig:mfa}-A3). The emergence of the slow patterns associated with sleep is appreciable already in the time domain (compare A1 to A3). The power spectrum of the EEG during task execution can be well fitted by a power law (\cref{fig:mfa}.B1), while rest and sleep EEG show significant deviations (\cref{fig:mfa}.B2–B3). In particular, EEG during rest is characterized by a strong alpha wave (\SIrange{8}{12}{\hertz}), manifested as a peak in the power spectrum in the alpha-band frequencies (\cref{fig:mfa}.B2). The EEG during sleep presents a predominance of lower frequencies in the delta (\SIrange{1}{4}{\hertz}) and theta (\SIrange{4}{8}{\hertz}) bands (\cref{fig:mfa}.B3), which characterize non-rapid eye movement sleep. Thus, both rest and sleep EEG examples present a clear deviation from a typical $1/f^a$ scale-invariant behaviour.

We thus adopted the multifractal framework described above to test whether it is possible to characterize these types of EEG under a more general definition of scale invariance. For each EEG sample, we computed the wavelet structure functions $S_x(j, q)$ and estimated the uniform H\"ol\-der exponent $H^\text{min}_x$ using Daubechies wavelets with 3 vanishing moments~\cite{daubechies1992ten} (the result confirms that  this choice yields sufficiently smooth wavelets to analyze such data). In \cref{fig:mfa}.C we show the structure functions $S_x(j, q)$ for $q = 1, 2$. For the EEG during task, $\log_2 S_x(j, q)$ can be well fitted by a linear function of $j$ (\cref{fig:mfa}.C1, dashed lines), making it possible to define a multifractal scaling via the scaling function $\eta_x(q)$. Note that we restricted the fit to the frequency range \SIrange{0.1}{50}{Hz} which contains the most relevant physiological information and is not affected by filtering and acquisition limitations (indicated by the shaded area in \cref{fig:mfa}.C). Both EEG during rest and sleep show significant deviations from linear behaviour of the structure functions in the log-log plots (\cref{fig:mfa}.C2–C3), which prevents possible estimation of the respective scaling functions. Similarly, while we can reasonably define a $H^\text{min}_x$ exponent for the EEG during active task execution (\cref{fig:mfa}.D1), the result is unreliable for the other two cases (\cref{fig:mfa}.D2–D3). 

To conclude, we report here that scale invariance of the EEG signal depends on the context of acquisition and is not a general feature. In particular, while scale invariance can be defined for EEG during mental activity, the same cannot be said of those brain states which are characterized by more regular rhythms and patterns such as sleep or rest. This result is not surprising as the EEG reflects processes occurring at various timescale (from milliseconds~\cite{hille1978} to minutes), generated by processes of different natures: ionic channels, neuronal spiking and bursting, transient changes of the membrane potential, or oscillations that reflect communication between different brain regions~\cite{buzsaki2006rhythms}. Depending on the context, the EEG signal can thus become dominated by patterns at specific timescales which break a possible underlying scale invariance. This observation justifies, in the context of EEG artifact correction to use wavelet algorithms such as WQN, which act on each scale independently.

\subsection{Wavelet coefficient distributions for EEG artifacts}

\begin{figure*}
    \includegraphics[width=\textwidth]{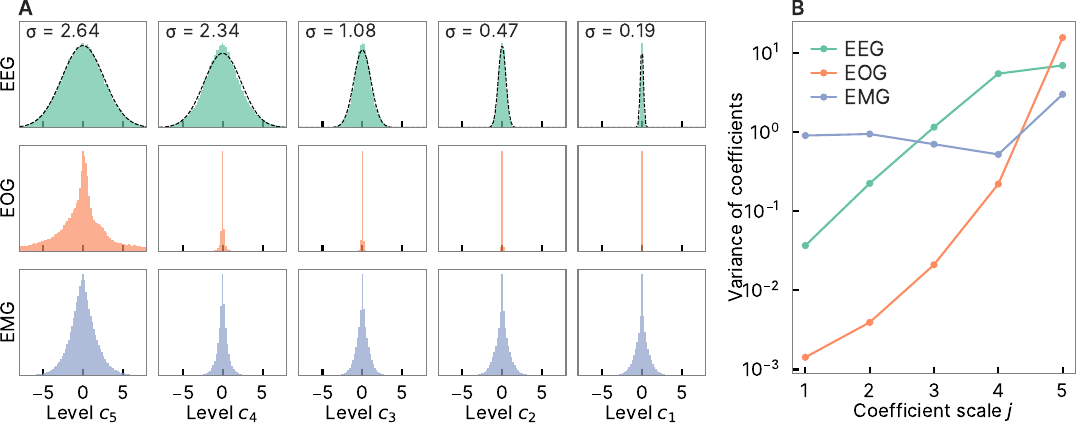}
    \caption{
        \textbf{Characterization of ocular and muscular artifacts compared to EEG in the wavelet domain.}
        \textbf{A.}~Distribution of the wavelet coefficients at different scales for EEG, EOG (ocular artifacts), and EMG (muscular artifacts). The EEG distributions can be fitted by Gaussian distribution with minor deviations (dashed line, standard deviation $\sigma$).
        \textbf{B.}~Plot of the variance versus scale. Variances for the pure EEG can be well approximated by a linear function (in log-log scale), compatibly with scale invariance shown in \cref{fig:mfa}, while artifacts show deviations.
    }
    \label{fig:coeff_dist}
\end{figure*}

\noindent We now characterize the EEG and artifactual signals by examining the distribution of their wavelet coefficients. We considered EEGs containing two common types of artifacts, ocular (EOG) and muscular (EMG), and compare them to uncontaminated EEGs. Ocular artifacts are caused by eye movements, which can alter the recorded electric potential since the eye acts as a dipole (with a difference of potential between the cornea and the fundus) while muscular artifacts are derived from the interfering electrical activity produced by muscle contraction. We decomposed two-second signals from the Denoise-Net dataset~\cite{zhang2020eegdenoisenet} using a Daubechies wavelet with two vanishing moments to obtain the coefficient distributions for each scale, shown in \cref{fig:coeff_dist}.A.

The distributions of wavelet coefficients for the uncontaminated EEG can be approximated by a normal distribution (\cref{fig:coeff_dist}.A, first row, dashed line) with minor deviations. In the case of ocular artifacts (\cref{fig:coeff_dist}.A, second row), the non-zero coefficients are concentrated at large scales (low frequency) with distinctive asymmetric distribution. Wavelet coefficients for muscular artifacts (\cref{fig:coeff_dist}.A, third row) show a deviation from the normal distribution, but are spread across multiple scales.  We investigate  a possible  scale invariant behaviour by plotting the variance of the wavelet coefficients versus the scale $j$ (\cref{fig:coeff_dist}.B), corresponding to the wavelet scaling function $S(j, q=2)$.

We notice that the variances of the uncontaminated EEGs can be approximated by a linear relation in log-log scale, thus showing some form of scale invariance, as shown in \cref{fig:mfa}. The variance of ocular artifacts decays rapidly with smaller scales, while muscular artifacts are characterized by an almost constant variance across all scales.

To conclude, we have shown that EEG artifacts present significant deviations from the uncontaminated EEG signal both in distribution of the wavelet coefficients and their persistence across scales. Moreover, significant differences in the scaling behaviour can be observed between distinct families of artifacts, such as ocular and muscular. 

\section{Properties of the WQN algorithm}\label{sec:properties}

\noindent We briefly recall the wavelet quantile normalization (WQN) algorithm~\cite{dora2022adaptive}. WQN is an adaptive method that allows to attenuate transient artifacts based on statistics estimated from clean regions of the signal. The algorithm was especially designed to be applied on EEG signals in the context of brain monitoring, where signal corruption is a consequence of artifacts generated by motion of electrodes, muscular activity or eye motion that alter the physiological signal generated by neuronal activity~\cite{tatum2011artifact,schomer2012niedermeyer}.

First, the signal is decomposed on a wavelet basis according to \cref{eq:signal_dwt}. We assume here that the artifacted intervals are well identified and isolated. They usually consist of a small portion of the total EEG signal. Once an artifact segment has been localized, it is associated to a clean reference segment where we expect the underlying signal to be similar but uncontaminated by artifacts. Note that these uncontaminated statistics depend on the patient and timing so that they cannot be acquired in advance.

\begin{figure*}
    \includegraphics[width=\textwidth]{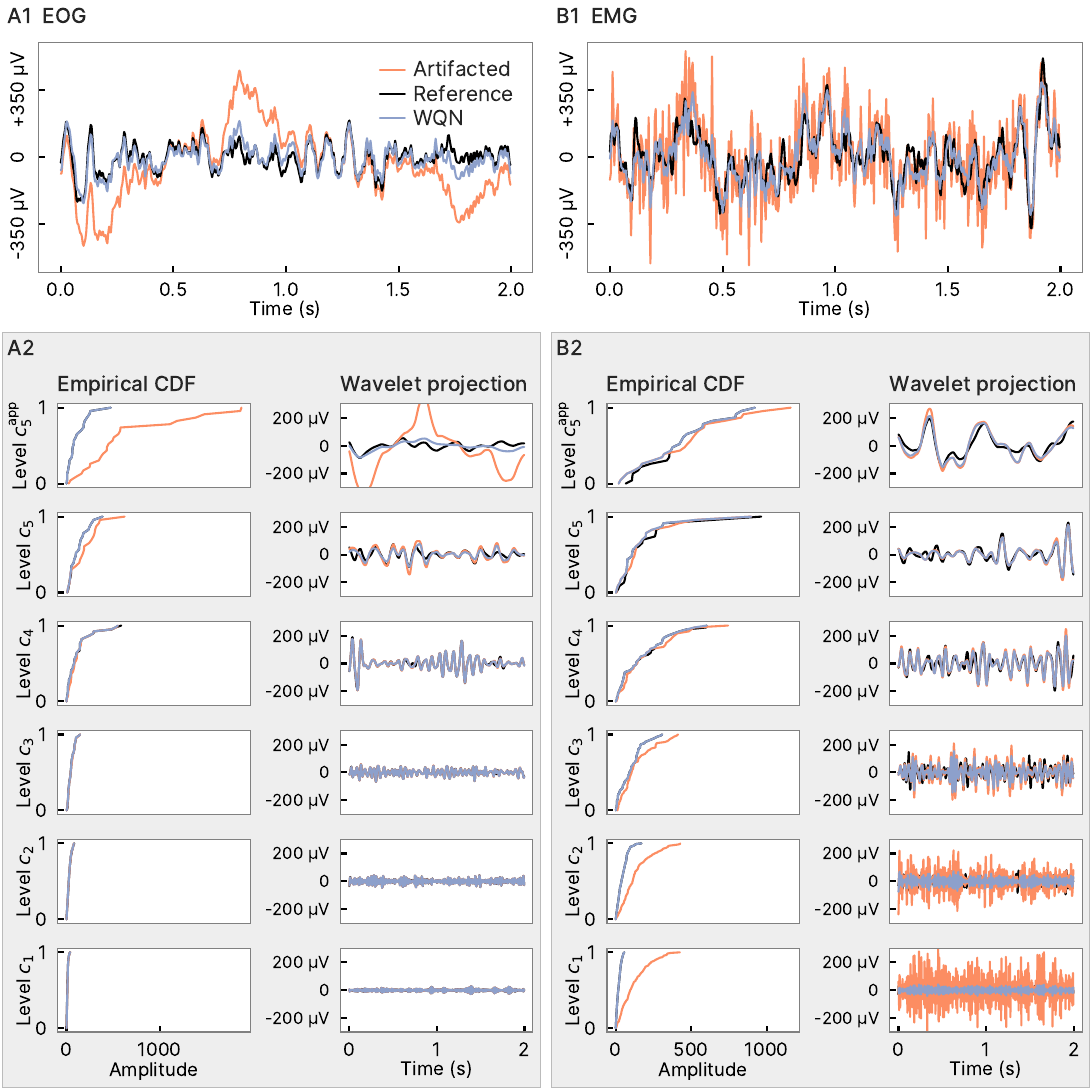}
    \caption{
        \textbf{Correction of EEG signals perturbed by ocular and muscular artifacts.}
        \textbf{A1.}~EEG signal corrupted by ocular artifact (EOG) and its reconstruction by WQN.
        \textbf{A2.}~Left: cumulative density function and its mapping from artifact (orange) to reference (black). Right: Projections on the wavelet basis showing attenuation of the artifactual components on different scales.
        \textbf{B1–B2.}~Similar presentation of WQN correction for an EEG signal contaminated by muscular artifact (EMG).
    }
    \label{fig:wqn_mapping}
\end{figure*}

In the original implementation, such reference intervals are defined by considering a temporally adjacent uncontaminated portion of the signal of roughly the same length as the contaminated portion. For each decomposition scale $j$, we define the wavelet coefficients associated with the artifacted and reference intervals by $c^{\text{(art)}}_{j}$ and $c^{\text{(ref)}}_{j}$ respectively. In practice, since we work with finite intervals, the wavelet decomposition is carried on up to scale $M$ (i.e. $j = 1, \dots, M$) guaranteeing the presence of a sufficient number of coefficients for the largest scale (e.g. $\text{Card} \left\{ c^\text{(art)}_{M,k} \right\} > 30$).\\
In the second step, the wavelet coefficients $c^\text{(art)}_j$ of the artifacted sample are modified in order to fit the statistics of the reference coefficients $c^\text{(ref)}_j$ at the corresponding scale $j$.  This normalization is performed, scale by scale, by computing the empirical cumulative density functions (CDF) $F^{\text{(ref)}}_{j}$, $F^{\text{(art)}}_{j}$ of the coefficients amplitude for artifacted and reference signal respectively, defined as
\begin{equation}
    F_{j}(x) = \frac{1}{N_j} \sum_{k = 1}^{N_j}  {1}_{\left| c_{j,k} \right| <\,x},
\end{equation}
where $N_j$ indicates the number of coefficients at scale $j$ and ${1}_{\left| c \right| <\,x}$ is the indicator function which takes value 1 if $\left| c \right| <\,x$ and 0 otherwise. The wavelet coefficients $c^\text{(art)}_j$ are then modified so that the distribution of their amplitude matches that of the reference segment, via the mapping $T_j$ defined as
\begin{equation}\label{eq:cdf_map}
    T_j(x) = F^{\text{(ref)}\, -1}_{j} \left( F^{\text{(art)}}_{j}(x) \right),
\end{equation}
where $F^{\text{(ref)} \, -1}_{j}$ indicates the generalized inverse of $F^{\text{(ref)}}_{j}$ (in the sense of completed graphs for discontinuous increasing functions) see \cite{dora2022wqn} and Fig. \ref{fig:wqn_mapping}. Finally, the normalization function
\begin{align}\label{eq:coeff_normalization}
    \lambda_{j}(c) = sign(c) \cdot \min\left\{\left| c \right|,\, \left|T_j \left(  c  \right) \right|\right\},
\end{align}
maps a coefficient $c$ from $c^{\text{(art)}}_j$ to its possibly attenuated value. The corrected coefficients are thus defined by
\begin{equation}\label{eq:corrected_coeffs}
    c^{\text{(corr)}}_{j,k} =  \lambda_j \left( c^{\text{(art)}}_{j,k} \right).
\end{equation}
\Cref{eq:coeff_normalization} ensures that the norm of wavelet coefficients is never increased, a key requirement to guarantee the regularity of the algorithm, as we will show in \cref{sec:regularity}.
The corrected version of the signal is obtained by replacing the artifacted coefficients $c^\text{(art)}_{j,k}$ by the corrected coefficients $c^\text{(corr)}_{j,k}$ and then inverting the discrete wavelet transform.

\subsection{Illustration of the WQN algorithm}

\noindent We present two examples to illustrate how the WQN algorithm works and how it adapts to artifacts affecting different scales. To this aim, we have contaminated an EEG signal by adding an ocular artifact (EOG) and a muscular artifact (EMG), as illustrated in \cref{fig:wqn_mapping} (A1--B1, first row). In \cref{fig:wqn_mapping}.A2-B2 (left) we represent how the wavelet coefficients are transported from the artifacted cumulative density function to the reference one. In the case of ocular artifact (\cref{fig:wqn_mapping}.A2) most of the transport occurs in the first scales (level 5), corresponding to low frequency components, while shorter scales (levels 1–5) which are not affected by the EOG artifact are left almost unmodified. The reconstruction of the signal, which can be seen as the sum of the corrected wavelet projections, is shown in \cref{fig:wqn_mapping}.A1 (blue). The energy of the EOG artifact is mostly concentrated in the low frequencies, corresponding to the approximation coefficients $c_5^{app}$ (the coarsest scale of the wavelet decomposition). Appropriately, most of the correction takes place at this scale (\cref{fig:wqn_mapping}.A2, first row).

In a second example (\cref{fig:wqn_mapping}.B) we present an EEG signal corrupted by a muscular artifact (EMG), which is characterized by a  wide frequency signature (as shown in \cref{fig:coeff_dist}), with high frequency components significantly more powerful than the EEG signal. In \cref{fig:wqn_mapping}.B2, we show how the WQN algorithm adapts to this different artifactual signature by attenuating the artifactual component in both shorter scales (levels 1–3) and larger scales (level 5). In conclusion, although the EMG and EOG artifacts are characterized by different statistics, the adaptive approach of the WQN algorithm makes it effective at reconstructing the original signal by performing appropriate corrections on a scale-by-scale basis.

\subsection{Regularity properties of the WQN algorithm}\label{sec:regularity}

\noindent In this section, we use the wavelet decomposition to derive a regularization property of the WQN algorithm. We recall that wavelets are unconditional bases of most classical function spaces \cite{Jaf9}, such as Sobolev spaces $H^{s,p}$ for $1< p < \infty$ or closely related Besov spaces $B^{s, q}_p$ for $0< p, q < \infty$ see e.g. \cite{Mey90I,jaffard2001wavelets}. This implies that these spaces have a wavelet characterization which bears on the moduli of the wavelet coefficients, and which is an increasing function of each of these moduli. For example, a function $f$ belongs to $B^{s, q}_p$ if its wavelet coefficients satisfy the condition
\begin{equation} \label{caracbes}
    \left(  \sum_n (2^{(1/p-1/2 -s)m} | c_{m,n}|)^p \right)^{1/p} \in l^q ,
\end{equation}
and \eqref{caracbes} yields a norm  (or a quasi-norm, when $p$ or $q$ is less than 1) which is equivalent to the Besov norm.\\
By construction, the WQN algorithm is { \sl wavelet decreasing} (i.e. it does not increase the size of the wavelet coefficients) and thus has the following regularity property:
For  $0< p, q \leq  \infty$, and for any $s \in \RR$,  it maps functions of a Besov or Sobolev space into the same space. Since the mapping is not linear, this property does not imply that it is continuous on the corresponding functional space; this question is relevant as the continuity of a denoising algorithm is a prerequisite to guarantee its numerical robustness.  In order to investigate this problem, it is useful to compare the present algorithm with the classical {\sl wavelet thresholding}  and { \sl wavelet shrinkage}  algorithms, which are conceptually simpler, and where the same problem arises.

\subsubsection{Comparison with wavelet thresholding and wavelet shrinkage}

\noindent Wavelet thresholding and wavelet shrinkage were introduced to perform denoising of signals or images without smoothing the signal to be recovered (in contradistinction with convolution-based techniques). In this context, the eliminated ``noise'' is assumed informally to be characterized with statistics of wavelet coefficients which, at a given scale,  are assumed to be stationary, of small amplitude and with short range correlations. The algorithm is efficient if the statistical properties of the signal to be recovered strongly differ from the noise i.e. if its wavelet coefficients form a \textit{sparse sequence} (most of them almost vanish), and the other coefficients are large. Note that this situation is opposite to the one we consider in the present article, where the artifacts to be eliminated have a sparse signature while the signal to be recovered presents the statistical properties of such a noise. However, wavelet thresholding and wavelet shrinkage  have also been used in such contexts: once the splitting has been performed, one keeps the ``noisy'' part instead of the sparse one \cite{inuso2007waveletica,sweeney2012artifact,chavez2018surrogate}. It is therefore legitimate to compare their performance with the WQN algorithm.\\
We briefly recall the wavelet thresholding. Given a threshold level $t >0$,  wavelet thresholding in a given wavelet basis is defined as follows: once an appropriate normalization of the wavelet coefficients has been chosen, the wavelet thresholding mapping is the operator $T$ which maps the wavelet coefficient $c_{m,n} $ to
\[ d_{m,n} = f_t(c_{m,n}) . \]
where
\begin{equation} \label{wavthr}
    f_t (x) =  x 1_{ [-t,t]} (x).
\end{equation}
The mapping $T$ is wavelet decreasing, and therefore maps functions in a Besov or Sobolev space to the same function space. Nonetheless, the function  $f_t$ is discontinuous and consequently, the operator $T$ is not continuous on any of these spaces. Indeed, to show this property, we consider two functions $f$ and $g$ with respectively wavelet coefficients $c^1_{m,n}$ and $c^2_{m,n}$ which coincide, except for one wavelet coefficient, such that these coefficients for $f$  and for $g$  respectively  are $t-\varepsilon$, and  $t+\varepsilon$.  Taking  $\varepsilon$  arbitrarily small, the Besov norm of $f-g$ can be made arbitrarily small, but the Besov norm of $T(f)- T(g)$ is  (up to the normalization factor of the corresponding wavelet coefficient) $t+\varepsilon$, and therefore does not tend to zero, when $\varepsilon \rightarrow 0$. This lack of continuity implies numerical instabilities of the algorithm which are well documented, see e.g.\cite{antoniadis2007wavelet,chambolle1998nonlinear,pang2023sparse}.\\
We now discuss the wavelet shrinkage algorithm, which is based on the function
\begin{equation}  \label{wavthr2}
    g_t (x) =  sgn (x) \cdot (|x| -t )^+.
\end{equation}
Once an appropriate normalization of the wavelet coefficients has been chosen, the wavelet shrinkage operator $U$ is defined as mapping the wavelet coefficient $c_{m,n} $ to
\[ e_{m,n} =  g_t(c_{m,n}) . \]\\
The mapping $U$ is wavelet decreasing,  and therefore maps functions in a Besov or Sobolev space to the same function space. But, additionally, in contradistinction with the previous case, $g_t$ is continuous, and therefore two functions $f$ and $g$ whose wavelet coefficients are close are now mapped to functions $U(f)$ and $U(g)$, which also have close wavelet coefficients. More precisely, since $g_t$ is Lipschitz, it follows easily that, for a given Besov or Sobolev space  $E$, $\parallel  U(f) -U(g) \parallel_E \leq C \parallel  f - g \parallel_E $, i.e. the mapping $U$ is Lipschitz in the corresponding Besov space. Refined stability properties of wavelet shrinkage can be attributed to $g_t$ continuity, i.e. (at least implicitly) to the continuity  of the mapping $U$, see \cite{antoniadis2007wavelet,chambolle1998nonlinear,pang2023sparse}.\\
One drawback of both of these algorithms is that they are \textit{local in the wavelet domain}, i.e. each wavelet coefficient is modified independently of the other ones, and therefore, they do not preserve the statistics of wavelet coefficients~\cite{dora2022wqn}. This phenomenon had no negative impact when wavelet thresholding and wavelet shrinkage were used for their initial purpose, i.e. to restore the ``sparse'' part of the signal, but it becomes a major drawback when it is used in the opposite direction of recovering the ``noisy'' component, in which case restoring the right statistics of the signal can be a major issue. For instance, where the eliminated artefact was localized, wavelet thresholding or shrinkage set the wavelet coefficients to zero, thus leading to inhomogeneities in the restored signal. One of the purposes of the WQN algorithm is to circumvent this drawback by restoring everywhere the correct anticipated statistics of wavelet coefficients \cite{dora2022adaptive}. As a consequence, it is not local in the wavelet domain. The value attributed to a coefficient depends on the entire statistic of coefficients at a given scale, and therefore the analysis of the regularity properties of the algorithm is more involved than for wavelet thresholding and wavelet shrinkage; nonetheless, a preliminary investigation of its main features will be performed  in subsection \ref{sec:contprop}.

\subsection{Robustness with respect to wavelet basis}

\begin{figure*}
    \includegraphics[width=\textwidth]{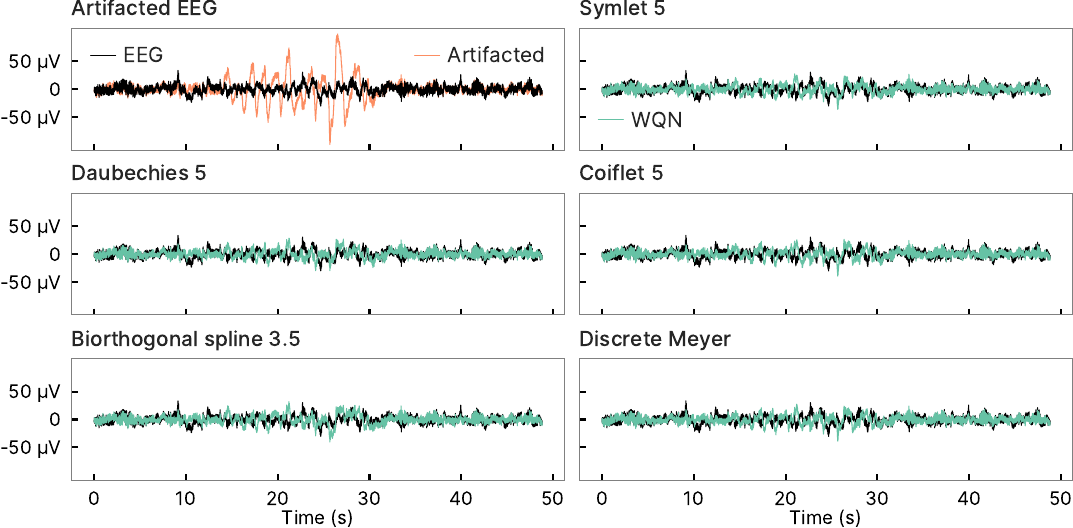}
    \caption{
        \textbf{Correction of electrode movement artifacts using the WQN algorithm. We show the time-plot of the corrections (green) of an added artifact (orange) using different wavelet bases: sym5, db5, coif3, bio3.5 and dmey \cite{mallat2009theory}.}
    }
    \label{fig:basis_independence}
\end{figure*}

To examine the performance of the WQN algorithm, we use five classical wavelet bases, currently used in signal processing (sym5, db5, coif3, bio3.5 and dmey)  \cite{mallat1989theory}. After we added an electrode moving artifact on an EEG signal, we use the WQN to remove the artifact (Fig. \ref{fig:basis_independence}), resulting in the green curves in the different sub-figures. To quantify this performance, we computed the  Average Root Mean Squared Error on the ensemble of data consisting in two types of artifacts (EOG ad EMG). We can conclude that the RMSE is quite independent of the wavelet bases with a mean around 0.03 for both types of artifacts (fig. \ref{fig:basis_independence}.A).\\
Finally, we tested the effect of increasing the number of vanishing moments of the wavelets. Again, we found that there was no noticeable  
consequence on the corrected artifact with the three following wavelet bases: Daubechies, Symlets and Coiflets. 
To conclude, the different wavelets bases have little influence on the corrected artifacts.

\subsection{Boundedness properties of the WQN algorithm in functional spaces}

\noindent As already mentioned above, since the WQN algorithm does not increase the size of the wavelet coefficients, it maps functions of a Besov or Sobolev space into the same space. One can also consider other function spaces which are based on histograms of wavelet coefficients at each scale and thus encapsulate more functional information than  Besov spaces do \cite{Jaf9,A06}. The maximal information which is invariant under the change of (smooth) wavelet basis, is encapsulated through the \textit{wavelet profile} of $f$, which is defined as follows. For a function $f$, we define
\[
    F_m (\alpha) = \text{Card} \left\{n: \;\; |c_{n,m} | \geq  2^{(\alpha + 1/2) m } \right\} ;
\]
and the wavelet profile $\nu_f (\alpha ) $  is
\[
    \nu_f (\alpha )  = \lim_{ \eps \rightarrow 0} \;
    \left[  \limsup_{j\rightarrow \infty} \left(\frac{\log (F_m(\alpha + \eps))}{\log (2^j) }\right) \right].
\]
This definition formalizes the following heuristic: There are about $2^{-\nu_f (\alpha )m}$ wavelet coefficients larger than $2^{(\alpha + 1/2) m}$. The corresponding function spaces are defined similarly:
Let $\nu (\alpha )$ be a nondecreasing function  which
takes values in $\{ -\infty \} \cup [0, 1]$. A function $f$ belongs to the
space $S^{\nu}$ if its wavelet coefficients satisfy. $ \forall \alpha \in \RR,$ $  \forall \eps >0, $
\[\forall
    C>0, \;
    \exists M
    \; \forall m \leq M \;\;\; \;
    F_m (\alpha ) \leq 2^{-(\nu (\alpha ) +\eps )m}. \]
Since the several wavelet algorithms that we considered are wavelet decreasing, it follows that  these algorithms map functions  which belong to a $S^{\nu}$ space to the same $S^{\nu}$ space.

\begin{figure*}
    \includegraphics[width=\textwidth]{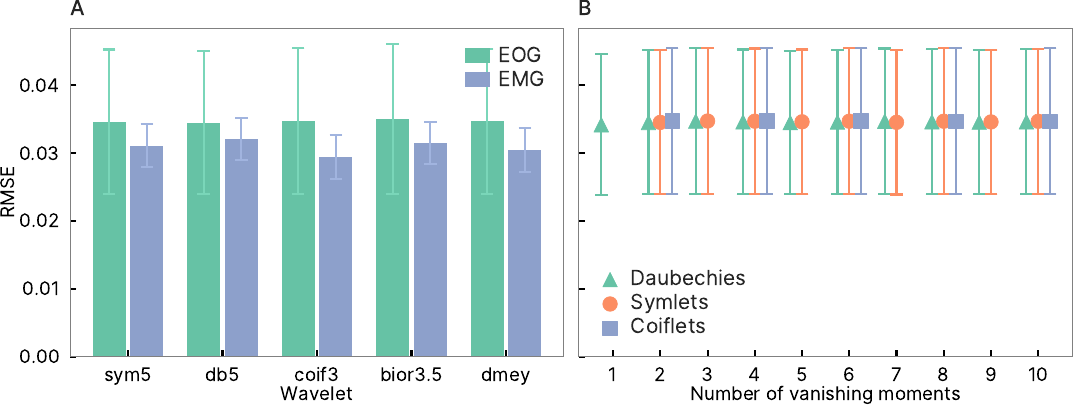}
    \caption{
        \textbf{WQN performance under different wavelet bases.}
        \textbf{A.}~Average Root Mean Squared Error (RMSE) of WQN for bootstrapped EEG signals contaminated by EOG and EMG, for the following wavelet bases: Symlet with 5 vanishing moments (sym5), Daubechies with 5 vanishing moments (db5), Coiflet with 6 vanishing moments (coif3), biorthogonal spline with 3 and 5 vanishing moments in synthesis and analysis wavelet respectively (bior3.5), discrete Meyer wavelet (dmey).
        \textbf{B.}~WQN performance for different number of vanishing moments, calculated on the EOG contaminated dataset.
    }
    \label{fig:basis_independence_b}
\end{figure*}

\subsection{Continuity properties}\label{sec:contprop}

\noindent After having considered the issue of the boundedness of the wavelet transport algorithm on several classes of function spaces, we now turn to the problem of its continuity.  We start by a simple remark which will allow to position the problem correctly. Assume that the unaltered data on which the histograms of wavelet coefficients are recorded has a given finite length $L$. Then one computes the number $N_m \sim L 2^{-m} $ of wavelet coefficients at scale $2^m$, which constitute the reference signal on which the altered data will be mapped.  The transport algorithm maps the wavelet coefficients of the altered signal on this finite set of cardinality $N_m$. Now assume that, for the altered signal, two wavelet coefficients of consecutive size are extremely close; if their sizes are exchanged,  the coefficients $c_{m,n}$ and $c_{m,l}$ on which they are mapped will also be exchanged;  it follows that, no matter how close the two starting functions are picked,  one wavelet coefficient of their image will differ by the value $c_{m,n}-c_{m,l}$; it follows that, strictly speaking,  the WQN algorithm is not continuous on any Besov or Sobolev space.  However, this theoretical argument does not necessarily constitute a drawback in applications if we make the assumption that the coefficient distribution   for the reference (unaltered) data is continuous and is sampled with enough precision; indeed, in practice consecutive wavelet coefficients will be mapped to very close  values, and the discontinuity of the mapping would be of no practical consequence, since its ``jumps'' would be of very small size. Note that this assumption is satisfied by the data we consider since the histograms of wavelet coefficients follows a generalized Gaussian distribution (Fig. \ref{fig:coeff_dist}).  At this point, another phenomenon has to be taken into account: even if the repartition function of coefficients for the reference data is continuous (so that the ``target'' coefficients are very close), it is still possible that the altered signal has a large number of coefficients which are close to each other, so that a very small perturbation in the size of coefficients would exchange two coefficients of very different ranks. This argument shows that a continuity result for the functional operator underlying the algorithm cannot follow from making only the assumption of a regularity of the target probability density function (PDF). However, though such situations yield mathematical counterexamples, they are not met in practice, since the data on which the algorithm is applied  also display smooth PDFs (Fig. \ref{fig:coeff_dist}),  and therefore do not exhibit large clusters of coefficients taking almost the same value.

\section{WQN algorithm on pathological examples}

\begin{figure*}
    \includegraphics[width=\linewidth]{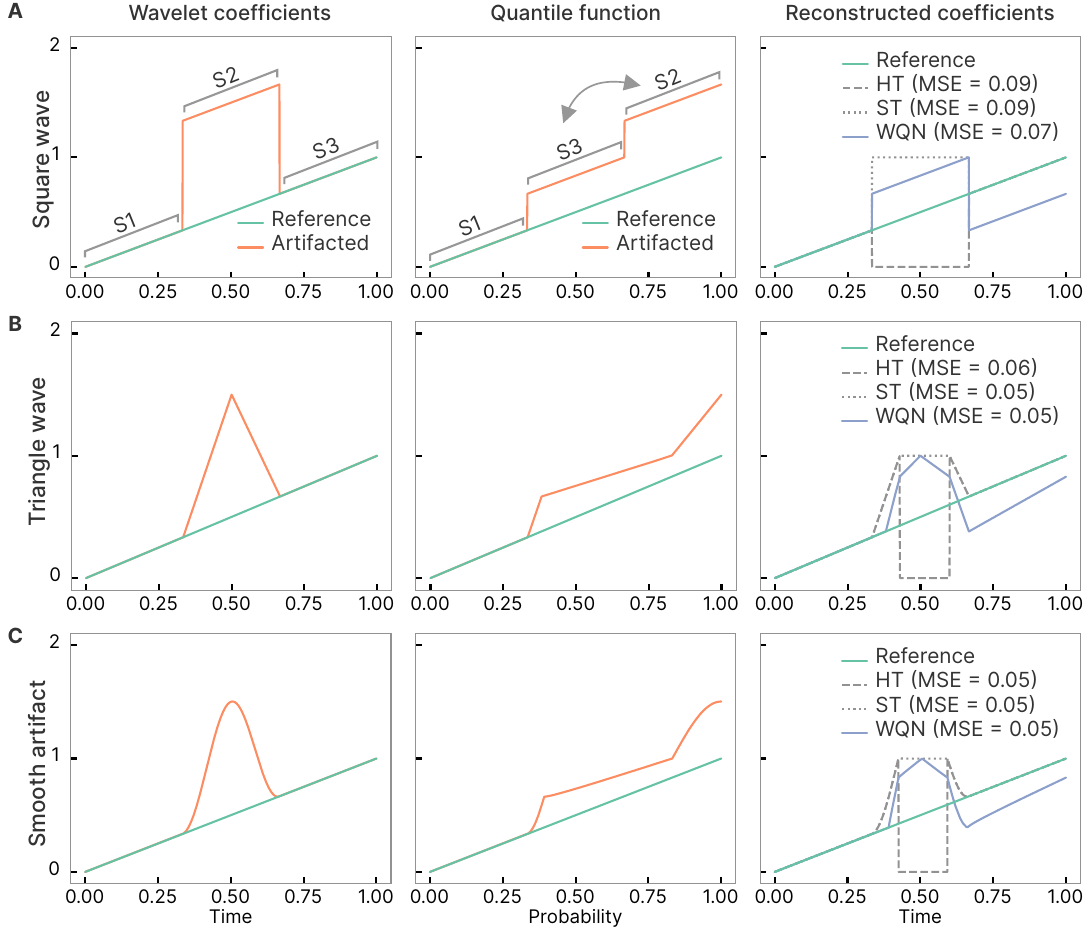}
    \caption{
        \textbf{Comparison of WQN and thresholding methods on pathological examples.}\,
        \textbf{A.}~Reference coefficients perturbed by additive square artifact $x_\text{sq}$. The WQN remapping causes swapping of coefficients, while thresholding methods act locally.
        \textbf{B.}~Signal with additive triangle perturbation $x_\text{tr}$.
        \textbf{C.}~Signal with a smooth perturbation $x_\text{cos}$. Non-local effects of WQN are similar to the triangle case.
    }
    \label{fig:pathological}
\end{figure*}

\noindent To further highlight the properties of the WQN algorithm, we present its application to pathological examples and compare its behaviour with soft (eq.~\ref{wavthr2}) and hard (eq.~\ref{wavthr}) thresholding.
We initially study how the algorithm operates on a single level of wavelet coefficients to highlight the non-local action of the WQN, which we previously discussed in \cref{sec:contprop}. We consider as a reference signal (in the wavelet space) a ramp $c_\text{ref}(t) = t$, as the simplest signal having a simple invertible CDF.
We then perturb the wavelet coefficients by adding three types of artifacts to the reference coefficients:
a square artifact $c_\text{sq}(t)=(H(t-t_1)-H(t-t_2))$ where $H$ is the Heaviside step function, a triangle 
\[
    c_\text{tr}(t)=
    \begin{dcases}
        \frac{2}{t_2 - t_1} t,     & \text{ if } \, t_1 \leq t \leq \frac{t_1 + t_2}{2},  \\
        1 - \frac{2}{t_2 - t_1}, & \text{ if }\, \frac{t_1 + t_2}{2} \leq t \leq t_2, \\
        0,                        & \text{ otherwise,}
    \end{dcases}
\]
and a cosine artifact
\[
    c_\text{cos}(t)=
    \begin{dcases}
         \frac{1 + \cos\left(\pi (2(t - t_1)/T - 1)\right)}{2},     & \text{ if } \, t_1 \leq t \leq t_2,  \\
        0,                        & \text{ otherwise,}
    \end{dcases}
\]
with $T = t_2 - t_1$.
Considering $c_\text{art} \in \{c_\text{sq},c_\text{tr}\, c_\text{cos}\}$, the final artifacted coefficients are given by
\begin{equation}
    c(t)=c_\text{ref}(t)+ c_\text{art}(t).
\end{equation}
In \cref{fig:pathological} we show the correction of coefficients $c(t)$ when applying the WQN quantile remapping and compare it with soft and hard thresholding. In the case of the square artifact (\cref{fig:pathological}.A), due to the non-locality of the WQN algorithm (see \cref{sec:contprop}), the mapping of the coefficients performed by WQN throught the inverse CDF (quantile function) can swap the values of group of coefficients (compare \cref{fig:pathological}.A first and second column). For wavelet thresholding, we consider the threshold to be equal to the maximum value of the unperturbed coefficients (which equals 1 in these examples). In the case of the square artifact, this results in perfect isolation of the artifacted coefficients, which are set to zero and 1 by hard and soft thresholding respectively. Interestingly, while the thresholding methods act locally, they do not result in a lower mean squared error with respect to WQN.
The cases of the triangle and cosine perturbation are presented in \cref{fig:pathological}.B and C respectively. In both cases, the WQN remapping still shows non-local swapping of coefficients, although the quantile function is continuous. Coefficients with the same amplitude exist in the unperturbed ($t < t_1, t > t_2$) and perturbed ($t_1 < t < t_2$) intervals, thus thresholding methods cannot perfectly isolate the perturbed coefficients. We note that the difference in smoothness between the triangle and cosine artifacts has no effect on the non-local action of the WQN algorithm, which produces similar reconstructions in the two cases. However, in both cases thresholding methods do not result in a lower mean squared error with respect to WQN.

\begin{figure*}
    \includegraphics[width=\linewidth]{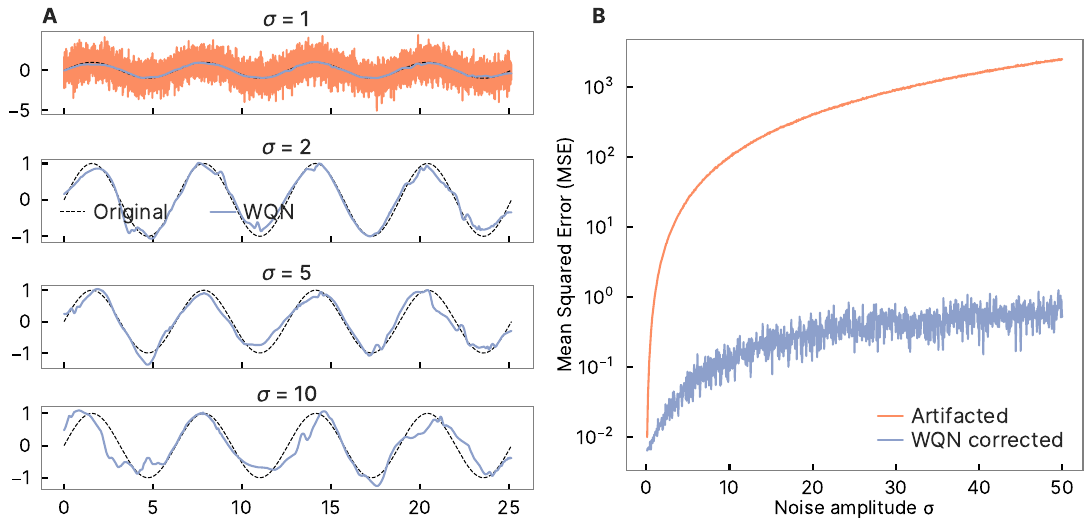}
    \caption{
        \textbf{WQN reconstruction of a sine wave perturbed by additive white noise.}\,
        \textbf{A.}~Reconstruction examples for $\sigma = 1, 2, 5, 10$.
        \textbf{B.}~Mean squared error (MSE) of the artifacted and the WQN-reconstructed signal with respect to the original sine wave, i.e.\ $\eE[(\sin (\omega t) - x(t))^2]$, plotted versus the artifactual noise amplitude $\sigma$.}
    \label{fig:white_noise}
\end{figure*}

As a final example, we evaluated how the WQN algorithm performs on a sine wave perturbed by white noise. This toy example was used to mimic a rhythmic brain activity perturbed by random artifact such as an epileptic seizure. The artifacted signal is given here as the sum
\begin{equation}
    x_\text{art}(t)= \sin (\omega t)+\sigma w(t),
\end{equation}
where $w$ is a centered Brownian noise of unit variance. In \cref{fig:white_noise}.A we show the signal reconstructed by WQN as we vary the noise amplitude $\sigma=1,2,5,10$. This example shows that WQN allows a robust reconstruction of rhythmic signals even in the presence of strong perturbations ($\sigma=10$). We quantify the noise reduction achieved by WQN in \cref{fig:white_noise}.B.

\section{Conclusion and final remarks}

\noindent We explored here the properties of EEG signals for different classes of brain states and artifacts. We presented evidence that, for particular states such as sleep and general anaesthesia, EEGs are not in general fully scale invariant;  the  deviations  are characterized by alternating dominance of specific frequency bands. Moreover, while the wavelet coefficients statistics at different scales are well approximated by Gaussian distributions in the case of clean EEG, those of common artifactual signals such as EOG and EMG are described by generalized Gaussians.

We then studied the properties of the WQN algorithm for EEG artifact removal. We showed how the WQN algorithm transports the wavelet coefficients from the distribution of an artifacted EEG signal into a distribution compatible with the uncontaminated EEG, allowing to restore the signal statistics. We found here that the WQN algorithm smoothens the discontinuities introduced by artifactual signals and provided insight into its regularity, continuity, and boundedness properties. We also highlighted how the remapping of wavelet coefficients through the quantile function can produce non-local effects, as opposed to traditional thresholding methods which always operate locally on the wavelet coefficients. Indeed, WQN can transport a wavelet coefficient to another position in time possibly located far away from the original position. This effect can thus generate local distortion and some information contained in the correlation structure of the wavelet coefficients can be lost. The classical functional space such as Besov spaces are unable to ``detect'' correlations between locations of wavelet coefficients, but other functional spaces, such as ``oscillations spaces'' can be more appropriate to detect them. In particular, they are not invariant under the ``shuffling'' of wavelet coefficients \cite{jaffard2004beyond}. Further investigations are needed in that direction.

Lastly, pathological cases can arise where the unperturbed physiological signal exhibits irregularities such as spikes and waves. For instance, this is particularly evident in certain forms of epilepsy or during deep anesthesia~\cite{rigouzzo2019eeg}. When an artifact occurs in correspondence of these dynamics, it would be interesting to test how WQN algorithm can map a singular signal into a reference that also contains singularities. It is of interest to evaluate the effectiveness of the WQN algorithm when an artifact occurs in correspondence of these dynamics, analyzing the ability of the WQN algorithm in effectively mapping a singular signal, affected by artifacts, to a reference signal that likewise exhibits singularities. Given its clinical relevance, such analysis warrants further exploration.

\section*{Code and data availability}

\noindent EEG, EMG, and EOG data used in the present article can be obtained from publicly available datasets~\cite{blankertz2007non,sweeney2012methodology,goldberger2000physiobank,quan1997sleep,zhang2020eegdenoisenet}. The Python code reproducing all results and figures is available on Zenodo (https://doi.org/10.5281/zenodo.8127712).

\section*{Competing interests}

\noindent The Authors declare no Competing Financial Interests.

\bibliographystyle{elsarticle-num}
\bibliography{bibliography}

\end{document}